%Paper: gr-qc/9508048
%From: SHORE@crnvma.cern.ch
%Date: Tue, 22 Aug 95 15:36:46 SET

\null

\input epsf

%*******************************************************************
%FONTS

\magnification 1200

%*******************************************************************
%FOOTNOTES - insert command \eightpoint to reduce

\newskip\ttglue

\font\eightrm=cmr8
\font\eighti=cmmi8
\font\eightsy=cmsy8
\font\eightbf=cmbx8
\font\eighttt=cmtt8
\font\eightsl=cmsl8
\font\eightit=cmti8
\font\sixrm=cmr6
\font\sixbf=cmbx6
\font\sixi=cmmi6
\font\sixsy=cmsy6

\def \eightpoint{\def\rm{\fam0\eightrm}% switch to 8-point type
\textfont0=\eightrm \scriptfont0=\sixrm \scriptscriptfont0=\fiverm
\textfont1=\eighti \scriptfont1=\sixi   \scriptscriptfont1=\fivei
\textfont2=\eightsy \scriptfont2=\sixsy   \scriptscriptfont2=\fivesy
\textfont3=\tenex \scriptfont3=\tenex   \scriptscriptfont3=\tenex
\textfont\itfam=\eightit  \def\it{\fam\itfam\eightit}%
\textfont\slfam=\eightsl  \def\sl{\fam\slfam\eightsl}%
\textfont\ttfam=\eighttt  \def\tt{\fam\ttfam\eighttt}%
\textfont\bffam=\eightbf  \scriptfont\bffam=\sixbf
 \scriptscriptfont\bffam=\fivebf  \def\bf{\fam\bffam\eightbf}%
\tt \ttglue=.5em plus.25em minus.15em
\setbox\strutbox=\hbox{\vrule height7pt depth2pt width0pt}%
\normalbaselineskip=9pt
\let\sc=\sixrm  \let\big=\eightbig  \normalbaselines\rm
}
%*******************************************************************
%GREEK LETTERS
\def\a{\alpha}

\def\d{\delta}
\def\e{\epsilon}
\def\h{\eta}

\def\l{\lambda}
\def\m{\mu}
\def\n{\nu}

\def\p{\pi}
\def\r{\rho}

\def\w{\omega}

\def\C{\Gamma}
\def\D{\Delta}

\def\S{\Sigma}

%*****************************************************************
%OTHER MACROS
\def\pl{\partial}

\def\dag{\dagger}

%*****************************************************************
%TITLE PAGE

{\nopagenumbers

\line{\hfil SWAT 95/71   }
\line{\hfil CERN-TH/95-229}
\vskip2cm
\centerline{{\bf ``FASTER THAN LIGHT'' PHOTONS AND ROTATING BLACK HOLES}}
\vskip1.5cm
\centerline{\bf R.D. Daniels{$^*$} and G.M. Shore{$^{*\dag}$} }
\vskip0.8cm
\centerline{$^*$ {\it Department of Physics}}
\centerline{\it University of Wales Swansea}
\centerline{\it Singleton Park}
\centerline{\it Swansea, SA2 8PP, U.K. }
\vskip0.8cm
\centerline{$^\dag$ {\it TH Division, CERN}}
\centerline{\it CH1211 Gen\`eve 23, Switzerland}
\vskip1.5cm
\noindent{\bf Abstract}
\vskip0.5cm
\noindent The effective action for QED in curved spacetime includes
equivalence principle violating interactions between the
electromagnetic field and the spacetime curvature. These interactions
admit the possibility of superluminal yet causal photon propagation
in gravitational fields. In this paper, we extend our analysis of
photon propagation in gravitational backgrounds to the Kerr
spacetime describing a rotating black hole. The results support
two general theorems -- a polarisation sum rule and a `horizon
theorem'. The implications for the stationary limit surface
bounding the ergosphere are also discussed.

\vskip2cm
\line{SWAT 95/71      \hfil}
\line{CERN-TH/95-229  \hfil}
\line{August 1995    \hfil}

\vfill\eject }

\pageno = 1

\noindent{\bf 1.~Introduction}
\vskip0.5cm
The possibility, originally discovered by Drummond and Hathrell
in 1980~[1], that photons propagating in curved spacetime may travel with
speeds exceeding the usual speed of light has been the subject of
renewed interest in the last two years[2-5]\footnote{$^*$}{\eightpoint
For earlier papers related to ref.[1] see
refs.[6,7]. Similar phenomena in non-gravitational backgrounds
are discussed in refs.[8-11,3]. }. The phenomenon arises
because vacuum polarisation in QED induces interactions between
the electromagnetic field and spacetime curvature. Such
interactions violate the strong principle of equivalence and allow
the possibility of spacelike photon propagation without necessarily
implying any violation of causality.

In their original paper[1], Drummond and Hathrell studied photon
propagation in the Schwarzschild spacetime, together with other examples
including gravitational wave and Robertson-Walker backgrounds.
In our previous paper[2], we extended this analysis to the
Reissner-Nordstr\"om geometry describing a charged black hole.
Here, we complete our survey of photon propagation in black hole
spacetimes by considering the Kerr geometry describing a
rotating black hole.

Our study is motivated by the hope that by examining special cases
with a particularly rich structure it may be possible to uncover
general properties of photon propagation in gravitational fields.
That hope is borne out by the results presented here.
In particular, it had been observed previously that the light cone
for radially directed photons in both the Schwarzschild and
Reissner-Nordstr\"om geometries remains unperturbed, i.e.~at $k^2=0$.
For the Kerr spacetime, we find that this is no longer true and photons
travelling on radial trajectories may indeed have velocities
differing from 1, either greater or smaller than the usual velocity
of light depending on the polarisation. However, we also find that
these corrections always vanish at the event horizon itself. The light
cone always remains $k^2=0$ at the horizon.

This result, together with a further observation that in all cases
the corrections to the photon velocity are equal and opposite for the
two transverse polarisations, was the motivation for ref.[12]
where they are formalised into two theorems -- a polarisation
sum rule and a `horizon theorem' (see section 5). This latter paper
also contains some related observations about electromagnetic
birefringence and the r\^ole of the conformal anomaly in
photon propagation.

\vskip0.7cm

\noindent{\bf 2.~Photon Propagation in Curved Spacetime}
\vskip0.5cm
In this paper, we consider the nature of photon propagation implied
by the action
$$
\C ~~=~~\int dx \sqrt{-g} \biggl(-{1\over4}F_{\m\n}F^{\m\n}
+ {1\over m^2}\Bigl( a R F_{\m\n} F^{\m\n} + b R_{\m\n} F^{\m\l}F^\n{}_\l
+ c R_{\m\n\l\r}F^{\m\n} F^{\l\r} \Bigr) \biggr)
\eqno(2.1)
$$
where the mass scale $m$ and constants $a$, $b$ and $c$ are regarded
as free parameters. This is the simplest action which incorporates
explicit equivalence principle violating interactions and leads to
possible modifications of the light cone. (Other more complicated
actions are considered in ref.[5].)

In fact, the expression (2.1) is generated as the effective action
in QED in curved spacetime including vacuum polarisation effects[1].
It is valid in the approximation of weak curvature and low frequency
photons, so that the neglect of higher powers of the curvature
tensor and extra covariant derivatives in the interaction terms
is justified. Roughly, (see ref.[13] for a careful discussion),
this effective action is a good approximation in the parameter
range $\l_c < \l < L$, where $\l$ is the photon wavelength,
$L$ is a typical curvature scale, and $\l_c = 1/m$ is the electron
Compton wavelength, $m$ being its mass. The electron provides the
appropriate quantum scale as the effect is generated by
vacuum polarisation diagrams involving an internal electron loop.
In this case, the coupling constants are given to one loop in terms of
the fine structure constant by $a = -{1\over144}{\a\over\p}$,
$b = {13\over360}{\a\over\p}$ and $c = -{1\over360}{\a\over\p}$.
Notice, however, that if this is the origin of the action (2.1),
then the validity of the results we deduce here concerning photon
propagation is only established for relatively low frequency
photons, whereas to consider the speed of real signal propagation and
the implications for causality we need to consider high frequencies.
The question of dispersion, the generalisation of the effective
action to include higher derivative terms and the physical
interpretation of the results will be discussed in ref.[13].

The Bianchi identity and equation of motion following from the
action (2.1) are\footnote{$^*$}{\eightpoint
See ref.[12] for an extended version of this section. }
$$
D_\m F_{\n\l} ~+~ D_\n F_{\l\m} ~+~ D_\l F_{\m\n} ~=~0
\eqno(2.2)
$$
and
$$
D_\m F^{\m\n} ~-~ {1\over m^2} \biggl( 2b R^\m{}_\l D_\m F^{\l\n}
{}~+~ 4c R^{\m\n}{}_{\l\r} D_\m F^{\l\r}\biggr) ~~=~~0
\eqno(2.3)
$$
where we have assumed that both the photon wavelength and $m^{-1}$
are small compared to the curvature scale.

The simplest way to determine the characteristics of photon propagation
from these equations is to use geometric optics. In the leading
geometric optics approximation, we write the electromagnetic field
strength as the product of a slowly varying amplitude and a rapidly
varying phase, i.e. $F_{\m\n} = f_{\m\n} \exp i\theta$, where
the wave vector is $k_\m = \pl_\m \theta$. In the quantum particle
interpretation, we identify $k_\m$ as the photon momentum.
The amplitude is constrained by the Bianchi identity to be of the
form $f_{\m\n} = k_\m a_\n - k_\n a_\m$, where the direction of
$a_\m$ specifies the polarisation. Light rays (photon trajectories)
are defined as the integral curves of the wave vector (photon
momentum). Without the additional term in eq.(2.3), these may easily be
shown to be null geodesics. This is no longer true when the
equivalence principle violating interactions are included,
nor is the light cone condition $k^2 = 0$ necessarily satisfied.

Our main concern is with the local modifications to the light
cone induced by these direct curvature interactions. Introducing
an orthonormal frame using vierbeins defined by $g_{\m\n} = \h_{ab}
e^a{}_\m e^b{}_\n$, where $\h_{ab}$ is the Minkowski metric,
the photon equation of motion (2.3) gives
$$
k^2 a_b ~-~ {2b\over m^2}\biggl(R_{ac}\Bigl(k^a k^c a_b - k^a k_b a^c
\Bigr)\biggr) ~-~ {8c\over m^2}\biggl(R_{abcd} k^a k^c a^d\biggr)
{}~~=~~0
\eqno(2.4)
$$
The polarisation vector is spacelike normalised, $a^b a_b = -1$,
and can be taken orthogonal to the momentum, $k^b a_b = 0$.
Of the three remaining degrees of freedom, in the quantum theory
only the two transverse polarisation states are physical.
Given a polarisation vector $a^b$ satisfying this equation, the
corresponding light cone condition is simply
$$
k^2 ~-~ {2b\over m^2} R_{ac} k^a k^c ~+~ {8c\over m^2}
R_{abcd} k^a k^c a^b a^d ~~=~~ 0
\eqno(2.5)
$$
We shall now explore the consequences of these equations for
the Kerr geometry.

\vskip0.7cm

\noindent{\bf 3.~The Kerr Spacetime}
\vskip0.5cm
The Kerr spacetime (see, e.g., refs.[14-16]) is described by the metric
$$
ds^2 ~~=~~ - \r^2 {\D\over \S^2} dt^2 ~+~ \r^2 {1\over \D} dr^2
{}~+~ \r^2 d\theta^2 ~+~ {1\over \r^2}\S^2 \sin^2\theta \bigl(
d\phi - \w dt\bigr)^2
\eqno(3.1)
$$
where
$$\eqalignno{
&\w = {2aMr\over \S^2} \hskip2cm       \r^2 = r^2 + a^2 \cos^2\theta \cr
&\D = r^2 - 2Mr + a^2  \hskip1cm \S^2 = (r^2+a^2)^2 - a^2\sin^2\theta \D
&(3.2) \cr}
$$
This metric represents the exterior spacetime of a rotating black hole.
It is axially symmetric about the rotation axis $\theta = 0$.
It is specified by two parameters, $M$ and $a$, where $M$ is the
mass and $Ma$ the angular momentum as measured from infinity.
For $a= 0$, it reduces to the Schwarzschild spacetime. In the
description below, we assume $a \le M$.

The condition $\D(r) = 0$, for which there is a coordinate singularity
similar to that in the Schwarzschild metric,
has two solutions, $r = r_{\pm} =
M \pm \sqrt{M^2 - a^2}$. The larger, $r=r_+$, is in fact the
event horizon, from within which no particle on a timelike or
null trajectory can escape to infinity. The region $r <  r_-$
contains a ring singularity.
Just as in the Schwarzschild geometry, there is a Killing vector
which is timelike in the asymptotic region (large $r$) but which
is spacelike within the event horizon. However, for the Kerr
spacetime, there is a further region outside the horizon,
$r_+ \le r <   r_E(\theta)$, for which the Killing vector remains
spacelike. This is the ergosphere. $r_E(\theta) = M +
\sqrt{M^2 - a^2 \cos^2\theta}$ is the value of $r$ for which
the metric component $g_{tt} = - \Bigl(1 - {2Mr\over \r^2}\Bigr)$
vanishes and the Killing vector is null. The outer limit of the
ergosphere, the stationary limit surface, is the inner boundary
of the region where particles travelling on timelike curves
can remain at rest relative to infinity. Within the ergosphere, even
null curves are pulled round in the direction of the rotation.
Particles may, however, escape to infinity from this region.
The situation is summarised pictorially in Fig.1, taken from ref.[15].
The stationary limit surface and event horizon coincide at the
poles, where the stationary limit surface becomes null.
At the equator, $r_E$ is equal to $2M$, the Schwarzschild radius.

We now introduce a local orthonormal frame. The appropriate
basis 1-forms are $e^a$ ($a=0,1,2,3$) with
$$
e^0 = e^0{}_t~dt \hskip1cm
e^1 = e^1{}_r~dr \hskip1cm
e^2 = e^2{}_\theta~d\theta \hskip1cm
e^3 = e^3{}_\phi~(d\phi - \omega dt)
\eqno(3.3)
$$
where the vierbeins are
$$
e^0{}_t = - \r {\sqrt \D\over \S} \hskip1cm
e^1{}_r = {\r\over \sqrt\D} \hskip1cm
e^2{}_\theta = \r \hskip1cm
e^3{}_t = -\w e^3{}_\phi = -{\w\over\r} \S \sin\theta
\eqno(3.4)
$$

The Kerr metric is Ricci flat, so $R_{ab} = 0$. There are six
independent non-vanishing components of the Riemann curvature
which we can choose in this frame to be[16]
$$\eqalignno{
&R_{0101} = A  \hskip1cm  R_{0202} = B  \hskip1cm  R_{0123} = C \cr
&R_{0231} = D  \hskip1cm  R_{0102} = E  \hskip1cm  R_{0113} = F
&(3.5) \cr}
$$
The complete set of non-vanishing components is
$$\eqalignno{
&R_{2323} = -R_{0101} \hskip1cm  R_{1313} = - R_{0202}
\hskip1cm  R_{1212} = - R_{0303} = R_{0101} + R_{0202} \cr
&R_{0312} = -R_{0123} - R_{0231}\hskip1cm
 R_{3132} = R_{0102} \hskip1cm
 R_{0223} = -R_{0113}
&(3.6)\cr }
$$
together with those related by the usual symmetries of $R_{abcd}$.
The quantities $A, \ldots , F$ are given by
$$\eqalignno{
&A~=~{Mr\over\r^6}\bigl(r^2 - 3a^2\cos^2\theta\bigr) {1\over\S^2}
\Bigl(2(r^2+a^2)^2 + a^2\D \sin^2\theta \Bigr) \cr
&B~=~-{Mr\over\r^6}\bigl(r^2 - 3a^2\cos^2\theta\bigr) {1\over\S^2}
\Bigl( (r^2+a^2)^2 + 2a^2\D \sin^2\theta \Bigr) \cr
 \Rightarrow ~~~&A+B~=~{Mr\over\r^6}\bigl(r^2-3a^2\cos^2\theta\bigr) \cr
&C~=~-{aM\cos\theta\over\r^6}
\bigl(3r^2 -  a^2\cos^2\theta\bigr) {1\over\S^2}
\Bigl(2(r^2+a^2)^2 + a^2\D \sin^2\theta \Bigr) \cr
&D~=~ {aM\cos\theta\over\r^6}
\bigl(3r^2 -  a^2\cos^2\theta\bigr) {1\over\S^2}
\Bigl( (r^2+a^2)^2 + 2a^2\D \sin^2\theta \Bigr) \cr
 \Rightarrow ~~~&C+D ~=~-{aM\cos\theta\over\r^6}\bigl(3r^2 -
a^2\cos^2\theta\bigr)  \cr
&E~=~ -{aM\cos\theta\over\r^6}\bigl(3r^2-a^2\cos^2\theta\bigr)
{3a\sqrt\D\over\S^2} (r^2+a^2)\sin\theta \cr
&F~=~ {Mr\over\r^6}\bigl(3r^2 - a^2\cos^2\theta\bigr)
{3a\sqrt\D\over\S^2}(r^2+a^2)\sin\theta
&(3.7) \cr }
$$
Introducing the notation $U^{01}_{ab} = \d^0_a \d^1_b - \d^0_b \d^1_a$
etc., we can rewrite the complete Riemann tensor compactly in the
following form:
$$\eqalignno{
R_{abcd} ~~=~~
&~~~~2A\Bigl(U^{01}_{ab}U^{01}_{cd}- U^{23}_{ab}U^{23}_{cd}
- U^{03}_{ab}U^{03}_{cd} + U^{12}_{ab}U^{12}_{cd} \Bigr) \cr
&+ 2B\Bigl(U^{02}U^{02} - U^{13}U^{13} - U^{03}U^{03} + U^{12}U^{12}
\Bigr) \cr
&+ C \Bigl(U^{01}U^{23} + U^{23}U^{01} - U^{03}U^{12} - U^{12}U^{03}
\Bigr) \cr
&+ D\Bigl(-U^{02}U^{13} - U^{13}U^{02} - U^{03}U^{12} - U^{12}U^{03}
\Bigr) \cr
&+ E \Bigl(U^{01}U^{02} + U^{02}U^{01} + U^{13}U^{23} + U^{23}U^{13}
\Bigr) \cr
&+ F \Bigl(U^{01}U^{13} + U^{13}U^{01} - U^{02}U^{23} - U^{23}U^{02}
\Bigr)
&(3.8) \cr}
$$
where we have suppressed the $a,b,c,d$ indices after the first line
for clarity. To obtain the curvature components $R_{\m\n\l\r}$ in the
coordinate frame $(t,r,\theta,\phi)$, simply replace $U^{01}_{ab}$
by $U^{01}_{\m\n} = e^0{}_\m e^1{}_\n - e^1{}_\n e^0{}_\m$ etc.
in this expression.

Notice the following simplifications for special cases.
On the event horizon, $\D(r_+) = 0$, so $E=F=0$, $A=-2B$ and $C=-2D$.
In the equatorial plane $\theta = \p/2$, $\cos\theta$ vanishes
and so $C=D=E=0$.

\vskip0.7cm

\noindent{\bf 4.~Photon Propagation in Kerr Spacetime}
\vskip0.5cm
We now return to eq.(2.4) describing photon propagation in curved
spacetime. For Ricci flat spacetimes such as Kerr, this reduces to
$$
k^2 a_b ~+~ \e R_{abcd} k^a k^c a^d ~~=~~0
\eqno(4.1)
$$
where we have written $\e = -8c/m^2$.
This is a set of three simultaneous linear equations for the
independent components of the polarisation $a_b$.
To solve these[1], it is convenient to introduce the following
linear combinations of momentum components:
$$
\ell_b = k^a U^{01}_{ab} \hskip1cm
m_b = k^a U^{02}_{ab} \hskip1cm
n_b = k^a U^{03}_{ab}
\eqno(4.2)
$$
together with the dependent combinations
$$\eqalignno{
&p_b = k^a U^{12}_{ab} = {1\over k^0} \bigl(k^1 m_b - k^2 \ell_b\bigr)\cr
&q_b = k^a U^{13}_{ab} = {1\over k^0} \bigl(k^1 n_b - k^3 \ell_b\bigr)\cr
&r_b =k^a U^{23}_{ab} = {1\over k^0} \bigl(k^2 n_b - k^3 m_b\bigr)
&(4.3) \cr }
$$
The vectors $\ell, m, n$ are independent and orthogonal to $k^a$.

We can therefore rewrite (4.1) as a set of equations for the
independent polarisation components $a.\ell$, $a.m$ and $a.n$
by contracting appropriately.
Substituting eq.(3.8) for the Riemann tensor, we therefore arrive
at the following set of equations:
$$\eqalignno{
0 ~~=~~k^2~a.\ell ~&+~ 2A\e \Bigl(\ell^2~a.\ell - \ell.r~a.r
- \ell.n~a.n + \ell.p~a.p \Bigr) \cr
&+~2B\e \Bigl(\ell.m~a.m - \ell.q~a.q - \ell.n~a.n + \ell.p~a.p
\Bigr)\cr
&+~C\e \Bigl(\ell^2~a.r + \ell.r~a.\ell - \ell.n~a.p - \ell.p~a.n
\Bigr) \cr
&+~D\e \Bigl(-\ell.m~a.q - \ell.q~a.m -\ell.n~a.p -\ell.p~a.n\Bigr) \cr
&+~E\e \Bigl(\ell^2~a.m + \ell.m~a.\ell + \ell.q~a.r + \ell.r~a.q
\Bigr) \cr
&+~F\e \Bigl(\ell^2~a.q + \ell.q~a.\ell - \ell.m~a.r - \ell.r~a.m
\Bigr)\cr
0 ~~=~~ k^2~a.m &+~~~ \ldots  \cr
0 ~~=~~ k^2~a.n~&+~~~ \ldots
&(4.4) \cr }
$$
To save space, we have only written the first in full. The second two
have a similar form and are easily reconstructed.

In principle, this set of equations could now be solved in general.
However, it is much more illuminating to look at a selection of
special cases which illustrate the most important features.
First, consider photon propagation in the equatorial plane.
\vskip0.7cm
\noindent (i)~~{\it Equatorial plane, radial motion}
\vskip0.5cm
To illustrate the method of solution, consider first radial motion
confined to the equatorial plane, where $C=D=E=0$.
The photon momentum components satisfy $k^2 = k^3 = 0$. The various
momentum-dependent terms appearing in eqs.(4.4) therefore simplify
considerably. In particular, we have $\ell^2 = k^0k^0 - k^1k^1$,~~
$m^2=n^2=k^0k^0$ and $\ell.m = \ell.n = m.n = 0$, while for the
others we find $p^2 = q^2 = k^1k^1$,~~$m.p = n.q = k^0k^1$, with
all other contractions vanishing.
For the polarisation projections, $a.p = k^1/k^0 a.m$,~~
$a.q = k^1/k^0 a.n$ and $a.r=0$.
Substituting these special results into eqs.(4.4)
and rewriting the system in matrix form, we find
$$\eqalignno{
&\left(\matrix{
 k^2 + 2A\e\ell^2
&0
&F\e \ell^2 {k^1\over k^0} \cr
 0
&k^2 + 2A\e k^1k^1 + 2B\e(k^0k^0+k^1k^1)
&0 \cr
 F\e k^0k^1
&0
&k^2 - 2A\e k^0k^0 - 2B\e(k^0k^0 + k^1k^1) \cr }\right) \cr
&{} \cr
&\times~~
\left(\matrix{ a.\ell \cr
 a.m \cr
 a.n \cr }\right)
{}~~=~~ 0
&(4.5) \cr }
$$

In general we would have to diagonalise to find the polarisation
eigenvectors, the corresponding values of $k^2$ being given as
solutions of the vanishing of the determinant. In this case, however,
there is a further simplification. We should regard $\e$ as a
small parameter (in fact, it is the dimensionless combinations
$A\e$, $B\e$ etc. which are small) since the original action
involving single powers of the curvature will be valid only
for $\l_c <   L$, and work to consistent order in small $\e$.
Since with $\e=0$ the light cone condition is just $k^2 = 0$,
in general we have $k^2 = O(\e)$. Now, for radial motion
$\ell^2 = -k^2$ so is of $O(\e)$, and therefore the off-diagonal
entry proportional to $F$ is actually of $O(\e^2)$ and should
be neglected at lowest order.

We therefore find the solutions:
\vskip0.2cm

\noindent $k^2 = 0$, ~~corresponding to the polarisation $a_a$
proportional to $\ell_a$,

\vskip0.2cm
\noindent $k^2 + 2A\e k^0k^0 + 2B\e (k^0k^0 + k^1k^1)$, ~~corresponding
to $a_a$ proportional to $m_a$,

\vskip0.2cm
\noindent $k^2 - 2A\e k^0k^0 - 2B\e (k^0k^0 + k^1k^1)$, ~~corresponding
to $a_a$ proportional to $n_a$.

\vskip0.2cm
\noindent The solution with $k^2 = 0$ corresponds to an unphysical
polarisation, $\ell_a = k^0 \d^1_a - k^1 \d^0_a$, while the two
physical transverse polarisations are proportional to
$m_a = k^0 \d^2_a$, i.e.~polarisation in the $\theta$ direction,
and $n_a = k^0 \d^3_a$, i.e. polarisation in the $\phi$ direction.
At this level the effect is non-dispersive and the (phase or group)
velocity shift is simply
$$
\d v ~~=~~ \biggl|{k^0\over k^1}\biggr| - 1
{}~~=~~ \pm (A+2B)\e ~~=~~ \mp \e \biggl({3Ma^2r^3 \over \r^6 \S^2 }
\D \biggr)
\eqno(4.6)
$$
for $\theta$ ($\phi$) polarisation respectively.

Two features of this result are immediately apparent. First, the
shift in $k^2$, or equivalently the velocity shifts away from the
conventional speed of light, are equal and opposite for the two
transverse polarisations. Second, the shift vanishes on the
event horizon, since $\D(r_+) = 0$. Both these observations turn out
to be examples of general theorems and we discuss them further in the
next section.

\vskip0.7cm
\noindent (ii)~~ {\it Equatorial plane, orbital motion}
\vskip0.5cm
Now consider photons travelling in the orbital ($\phi$) direction
in the equatorial plane. In this case, $k^1 = k^2 = 0$. The analysis
goes through in the same way as described above and gives the
solutions:

\vskip0.2cm
\noindent $k^2 + 2A\e k^0k^0 - 2B\e k^3k^3 - 2F\e k^0k^3 = 0$, ~~
corresponding to $a_a$ proportional to $\ell_a$,

\vskip0.2cm
\noindent $k^2 - 2A\e k^3k^3 + 2B\e k^0 k^0 + 2F\e k^0k^3 = 0$, ~~
corresponding to $a_a$ proportional to $m_a$,

\vskip0.2cm
\noindent $k^2 - (2A+2B)\e k^0k^0 = 0$, ~~
corresponding to $a_a$ proportional to $n_a$.

\vskip0.2cm
The two physical transverse polarisations, in the radial and $\theta$
directions, are given by the first two of these solutions respectively.
Notice, however, that now the light cone condition is not quadratic
in each of the momentum components separately due to the presence
of the $F$ term. This introduces a splitting in the velocities for
propagation with $k^3 > 0$ and $k^3 < 0$. For the radial polarisation
we find
$$
\d v ~~=~~(A- B \mp F) \e ~~=~~{3Mr^3\over \r^6 \S^2}
\Bigl( (r^2 + a^2)^2 + a^2\D \mp 3a(r^2 + a^2) \sqrt{\D} \Bigr)
\eqno(4.7)
$$
depending on whether the motion is with or against the direction
of spin. For the $\theta$ polarisation, $\d v$ has the opposite sign.
This expression simplifies on the horizon, where we have
$$
\d v |_{\rm horizon} ~~=~~ {3M\over r_+^3}~\e
\eqno(4.8)
$$
and on the stationary limit surface $r_E = 2M$, where (for $k^3 \le 0$)
$$
\d v|_{\rm stat.~lim.} ~~=~~ {3\over 2}{1\over r_E^4(r_E^2+2a^2)}
\Bigl(r_E^4 + 5a^2(r_E^2+a^2)\Bigr)~\e
\eqno(4.9)
$$

\vskip0.7cm
\noindent (iii)~~ {\it Radial motion, arbitrary direction}
\vskip0.5cm
In the Schwarzschild and Reissner-Nordstr\"om spacetimes, the light cone
for radial photons remains $k^2 = 0$ for any direction
(the solutions are of course spherically symmetric)
and for any value of the radial coordinate $r$. In the Kerr spacetime,
however, we have just seen that $k^2 \ne 0$ for radial photons in the
equatorial plane except at the horizon $r = r_+$.
We now want to check whether this result remains true independent
of the angle $\theta$ to the polar axis at which the photons are
directed.

The calculation follows the lines of (i) except that now the curvature
components $C$, $D$ and $E$ are non-vanishing. Eq.(4.5) generalises
to:
$$\eqalignno{
&\left(\matrix{ k^2 + 2A\e\ell^2
&E\e \ell^2
&F\e \ell^2 {k^1\over k^0} \cr
E\e k^0k^0
&k^2 + 2A\e k^1k^1 + 2B\e (k^0k^0 + k^1k^1)
&-(C+2D)\e k^0k^1 \cr
F\e k^0k^1
&-(C+2D)\e k^0k^1
&k^2 -2A\e k^0k^0 -2B\e (k^0k^0 + k^1k^1) \cr }\right)\cr
&{} \cr
&\times
\left(\matrix{a.\ell \cr
a.m \cr
a.n \cr }\right)
{}~~=~~ 0
&(4.10) \cr }
$$
Again there is a solution $k^2 = 0$ corresponding to the unphysical
polarisation $a_a$ proportional to $\ell_a$.
The transverse polarisations diagonalising (4.10) are complicated
$r$ and $\theta$ dependent linear combinations of the unit vectors in
the $\theta$ and $\phi$ directions, the corresponding velocity
shifts being
$$\eqalignno{
\d v ~~&=~~ \pm \Bigl(~(A+2B)^2 + {1\over 4}(C+2D)^2~\Bigr)^{1\over2} \cr
&=~~ \pm \e \D {3Ma^2\sin^2\theta\over \r^6 \S^2}
\Bigl(r^2(r^2-3a^2\cos^2\theta)^2 + {1\over4}a^2\cos^2\theta(3r^2
- a^2\cos^2\theta)^2 \Bigr)^{1\over2}
&(4.11) \cr }
$$
Notice that $\d v = 0$ along the polar axis, while of course the previous
result is recovered in the equatorial plane. Most important, however,
we again see that the velocity shifts are equal and opposite for
the transverse polarisations and that, independently of $\theta$,
$\d v = 0$ on the event horizon where $\D(r_+)$ vanishes.

\vskip0.7cm

\noindent{\bf 5. The Polarisation Sum Rule, the Horizon and the
Stationary Limit Surface}
\vskip0.5cm
These results on modifications of the light cone in special cases,
together with those previously obtained in refs[1,2], motivated
the formulation of two general theorems which were stated precisely
and proved in ref.[12] (see also ref.[17]). These are:
\vskip0.5cm
\noindent{\it Polarisation Sum Rule}
\vskip0.3cm
\noindent This relates the sum over the transverse polarisation states
to an appropriate projection of the Ricci tensor, viz.
$$
\sum_{\rm pol} k^2 ~~=~~ -{1\over m^2} (4b+8c) R_{ac} k^ak^c
\eqno(5.1)
$$
\vskip0.5cm
\noindent{\it Horizon Theorem}
\vskip0.3cm
\noindent This states that at the event horizon, the light cone
for photons with momentum directed normal to the horizon remains
$k^2 = 0$, independent of the polarisation.

\vskip0.7cm
Clearly, the specialisation of the polarisation sum rule to Ricci flat
spacetimes, i.e. $\sum_{\rm pol} k^2 = 0$, or equivalently
$\sum_{\rm pol} \d v = 0$, is satisfied by all the examples in Kerr
spacetime discussed in section 4.

Similarly, for radial photons, eq.(4.12) ensures that $k^2 = 0$,
or equivalently $v = 1$, at the event horizon even though the
velocity differs from 1 for all other values of the radial coordinate.
The horizon theorem therefore ensures that the geometric event
horizon remains a true horizon for photons propagating according to the
action (2.1), even in a spacetime with as rich a structure as Kerr.

In the light of this, it is interesting to ask whether the geometric
stationary limit surface $r = r_E(\theta)$ specified by $g_{tt} = 0$
retains its defining property for real photon propagation.
Recall that this is the surface within which even light signals
emitted against the direction of rotation are pulled round, through the
phenomenon of dragging of inertial frames[14,15], so that as measured by
an asymptotic observer they propagate in the direction of rotation
of the black hole.

The results of section 4(ii) show that this is not true. The fact that
the light cone is modified for photons emitted in the negative $\phi$
direction (see eqs.(4.7)(4.9)) even on the stationary limit surface
shows that the effective stationary limit surface is shifted from
$r_E(\theta)$ to a larger or smaller value of $r$ depending on the photon
polarisation.

The shift is most readily found using a trick used in ref.[1]
to calculate the modification to the bending of light in a
Schwarzschild spacetime. As we have seen, the light cone condition
is modified to $\tilde \h_{ab} k^ak^b = 0$ where $\tilde \h_{ab}$
differs from the orthonormal (Minkowski) metric by the terms of
$O(\e)$ calculated in section 4. Propagation with this modified
light cone in a spacetime with metric $g_{\m\n}$ is therefore
equivalent to conventional propagation with light cone $\h_{ab}k^ak^b=0$
in a modified geometry with metric $\tilde g_{\m\n} = \tilde \h_{ab}
e^a{}_\m e^b{}_\n$.

In this case, we have determined $\tilde \h_{ab}$ for propagation
in the counter-orbital direction in the equatorial plane (section 4(ii)).
Since $g_{tt} = - e^0{}_t e^0{}_t + e^3{}_t e^3{}_t$, we see
that evaluated on the (unperturbed) stationary limit surface
$e^0{}_t = e^3{}_t$. We therefore have immediately that
$$
\tilde g_{tt} ~=~ g_{tt} + 2\e (A-B+F) e^0{}_t e^0{}_t
\biggl|_{\theta = {\p\over 2}, r=2M}
\eqno(5.2)
$$
Setting $\tilde g_{tt} = 0$ gives the effective stationary limit
surface. In the equatorial plane, we therefore find
$$
r_E{}^{\rm eff} ~~=~~ 2M \pm \e (A-B+F) {a^2\over a^2 + 2M^2}
\eqno(5.3)
$$
depending on the polarisation,
the magnitude of the velocity shift term being given in eq.(4.9).

Finally, we should emphasise again that these results follow from
taking the action (2.1) literally. If instead we regard it as the
lowest-order effective action for QED, we should ask to what extent
the inherent approximations can be relaxed. The weak curvature
($\l_c < L$) approximation is difficult to improve on, so the
magnitude of the results will necessarily be extremely small in
the domain where the derivation is reliable. This limits the interest of
these results for astrophysical black holes. This is, however, just
the usual situation for quantum field phenomena in curved spacetime
such as, e.g., the Hawking effect[18]. The predicted effects are of
the order of a quantum scale divided by a curvature scale, so are
expected to become large only for microscopic black holes or in the
very early universe.
A more serious limitation of the action (2.1) is the implicit
restriction to relatively low-frequency photons ($\l_c < \l$),
which poses serious questions as to the observability and relevance
for signal propagation and causality of the shifts in the light cone
considered here. These issues will be addressed in a forthcoming
paper[13].
Despite the need for caution, however, the general theorems inferred
from our study of photon propagation in black hole spacetimes may well
prove to be valid outside these limitations. In particular,
the horizon theorem, whose proof relied on the physical result
that classically no gravitational radiation (or matter) crosses
the horizon[19], looks sufficiently robust to conjecture that it may
be true in general.

\vskip0.7cm
\noindent{\bf Acknowledgements}
\vskip0.5cm
We would like to thank Warren Perkins for useful discussions.
One of us (GMS) is grateful to Gabriele Veneziano and TH Division,
CERN for their hospitality while this paper was being completed.

\vfill\eject

\noindent{\bf References}
\vskip0.5cm

\settabs\+\ [&1] &G.M.Shore  \cr

\+\ [&1] &I.T. Drummond and S.J. Hathrell, Phys. Rev. D22 (1980)
343 \cr
\+\ [&2] &R.D. Daniels and G.M. Shore, Nucl. Phys. B425 (1994) 634 \cr
\+\ [&3] &J.L. Latorre, P. Pascual and R. Tarrach,
Nucl. Phys. B437 (1995) 60 \cr
\+\ [&4] &I.B. Khriplovich, Phys. Lett. B346 (1995) 251 \cr
\+\ [&5] &R. Lafrance and R.C. Myers, Phys. Rev. D51 (1995) 2584 \cr
\+\ [&6] &Y. Ohkuwa, Prog. Theor. Phys. 65 (1981) 1058 \cr
\+\ [&7] &A.D. Dolgov and I.B. Khriplovich, Sov. Phys. JETP
58 (1983) 671  \cr
\+\ [&8] &K. Scharnhorst, Phys. Lett. B236 (1990) 354 \cr
\+\ [&9] &G. Barton, Phys. Lett. B237 (1990) 559 \cr
\+ [1&0] &G. Barton and K. Scharnhorst, J. Phys. A26 (1993) 2037\cr
\+ [1&1] &S. Ben-Menahem, Phys. Lett. B250 (1990) 133 \cr
\+ [1&2] &G.M. Shore, `Faster than light photons in gravitational
fields -- \cr
\+\ &{}  &\hskip1cm causality, anomalies and horizons',
Swansea preprint SWAT 95/70 \cr
\+ [1&3] &R.D. Daniels and G.M. Shore, `Faster than light photons
in gravitational fields -- \cr
\+\ &{}  &\hskip1cm dispersion and the effective action',
Swansea preprint SWAT 95/55, in prep. \cr
\+ [1&4] &C.W. Misner, K.S. Thorne and J.A. Wheeler,
`Gravitation', Freeman, 1973 \cr
\+ [1&5] &S.W. Hawking and G.F.R. Ellis, `The large scale structure
of space-time', \cr
\+\ &{}  &\hskip1cm Cambridge University Press, 1973 \cr
\+ [1&6] &S. Chandrasekhar, `The mathematical theory of black holes',\cr
\+\ &{}  &\hskip1cm Oxford University Press, 1983 \cr
\+ [1&7] &G.W. Gibbons and M.J. Perry, to be published \cr
\+ [1&8] &S.W. Hawking,  Comm. Math. Phys. 43 (1975) 199 \cr
\+ [1&9] &S.W. Hawking, `The event horizon', 1972 Les Houches
lectures, ed. B. de Witt, \cr
\+\ &{}  &\hskip1cm Gordon and Breach, 1972 \cr

\vfill\eject

\vskip2cm

\centerline{\epsfbox{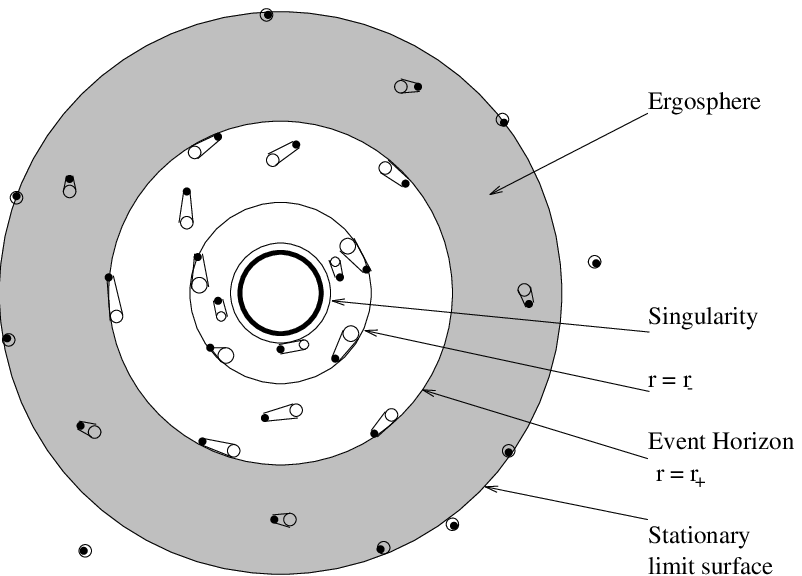}}

\vskip1cm

\noindent{\bf Fig.~1} ~~~A picture of the equatorial plane of the
Kerr geometry with $a^2<M^2$.
The circles represent the position after a short time interval
of photons emitted in the $r,\phi$ plane from the points represented
by the heavy dots. The distinctive `frame-dragging' effect giving
rise to the ergosphere is readily seen, as is the property of the
horizon that it cannot be crossed even by null trajectories.

\bye